\documentclass[useAMS,usenatbib]{mn2e}
\usepackage{url,ulem,times,graphicx,amsmath,amsfonts,amssymb,color,epsfig,epstopdf}
\usepackage{graphicx}
\usepackage{epsf}
\usepackage{color}
\usepackage{rotating}
\def\lsim{\mathrel{\lower0.6ex\hbox{$\buildrel {\textstyle <}
 \over {\scriptstyle \sim}$}}}
\def\gsim{\mathrel{\lower0.6ex\hbox{$\buildrel {\textstyle >}
 \over {\scriptstyle \sim}$}}}

\begin{document}

\title[Haloes in the cosmic web]{The cosmic web and the orientation of angular momenta}
\author[Libeskind et al.] 
{Noam I Libeskind$^1$, Yehuda Hoffman$^2$, Alexander Knebe$^3$, Matthias Steinmetz$^1$, \newauthor Stefan Gottl\"ober$^1$,   Ofer Metuki$^2$, \& Gustavo Yepes$^3$\\
 $^1$Leibniz-Institute f\"ur Astrophysik Potsdam (AIP), An der Sternwarte 16, D-14482 Potsdam, Germany\\
  $^2$Racah Institute of Physics, Hebrew University, Jerusalem 91904, Israel\\
  $^3$Grupo de Astrof\'\i sica, Departamento de Fisica Teorica, Modulo C-8, Universidad Aut\'onoma de Madrid, Cantoblanco E-280049, Spain
 }


\maketitle \begin{abstract} \vspace{1pt}
We use a 64$h^{-1}$Mpc dark matter (DM) only cosmological simulation to examine the large scale orientation of haloes and substructures with respect the cosmic web. A web classification scheme based on the velocity shear tensor is used to assign to each halo in the simulation a web type: knot, filament, sheet or void. Using $\sim 10^{6}$ haloes that span $\sim$3 orders of magnitude in mass the orientation of the halo's spin and the orbital angular momentum of subhaloes with respect to the eigenvectors of the shear tensor is examined. We find that the orbital angular momentum of subhaloes tends to align with the intermediate eigenvector of the velocity shear tensor for all haloes in knots, filaments and sheets. This result indicates that the kinematics of substructures located deep within the virialized regions of a halo is determined by its infall which in turn is determined by  the large scale velocity shear, a surprising result given the virilaized nature of haloes. The non-random nature of subhalo accretion is thus imprinted on the angular momentum measured at $z=0$. We also find that haloes' spin axis is aligned with the third eigenvector of the velocity shear tensor in filaments and sheets: the halo spin axis points along filaments and lies in the plane of cosmic sheets.
\end{abstract}

\section{Introduction}
\label{introduction} 

Satellite galaxies represent a particularly fascinating subset of the low redshift universe. Among their interesting properties is the so-called ``Holmberg effect'' \citep{Holmberg1969,Zaritskyetal1997}  which finds that satellites tend to be oriented close to the poles of their host disc galaxies, avoiding the planar regions. This type of polar alignment is also seen in the Milky Way \citep{KroupaTheisBoily2005,MetzKroupaJerjen2009} and was probably first noted over 3 decades ago \citep{LyndenBell1982}, an adequate explanation of which is still lacking.

External galaxies (for example in the SDSS) often exhibit a spatial alignment of satellites with the long axis of the host \citep{SalesLambas2004,Brainerd2005,Yangetal2006,AgustssonBrainerd2010} although this alignment is strongest for the older redder satellites - younger blue satellite galaxies exhibit no significant alignment. Its not clear whether the Milky Way's polar alignment is at odds with these results, as the surface brightness of many of the Milky Way's ultra faint satellite galaxies is many orders of magnitude lower than what can be observed extra-galactically.

In addition to their anisotropic spatial distribution at least 5 of the Milky Way's satellites (the Large and Small Magellanic Clouds, Ursa Minor, Carina and Fornax) have proper motions consistent with ``rotational support'' of the so-called ``disc-of-satellites'' \citep{MetzKroupaLibeskind2008}, an observation recently attributed to the angular momentum bias of satellites falling in along their hosts main axis \citep{Deasonetal2011}. Although examining a different environment and mass range of haloes, \cite{Tormen1997} was the first to suggest that galaxies fall into clusters anisotropically. Further to their work, the idea that satellites are preferentially (not randomly) accreted and retain a memory of the large scales from which they came has also recently been suggested \citep{Knebeetal2004,Libeskindetal2011a,Vera-Ciroetal2011}.

On larger scales a number of observational studies have hinted that galaxies separated by as much as $\sim 30$~Mpc (or 100s of times their virial radii) may also exhibit alignments albeit with the large scale distribution of matter \citep{hirataSeljak2004,Hirataetal2007,Mandelbaumetal2006,Okamuraetal2009}. On these scales, matter clusters into a ``cosmic web'' defined either geometrically \citep[e.g.][]{Sousbieetal2008} or dynamically \citep[e.g.][]{Hahnetal2007b,Forero-Romero2009}. 

Yet what lacks from the above studies is a clear link between the large scale cosmic web, the orientation of central galaxies and the preferential bias in satellite entry points, angular momenta and $z=0$ position. Some studies, such as \cite{Aragon-Calvoetal2007} and \cite{Hahnetal2007a} have used methods based on the tidal field or the Hessian of the density field to correlate intrinsic halo properties with the cosmic web. \cite{Aragon-Calvoetal2007} found a mass dependence for the (perpendicular) correlation between halo spins and cosmic  structure. \cite{Hahnetal2007a} examined how any correlation varied with redshift, extending their work in \cite{Hahnetal2010}, where they attempted to establish a link between central galactic discs and the cosmic web. In that paper they found an alignment between the spin of gaseous discs and the intermediate axis of the of the large scale tidal field, consistent with linear tidal torque theory \citep[e.g.][]{NavarroAbadiSteinmetz04, LeeErdogdu2007}. However none of these studies examined the orbits of satellite galaxies.

In this \textit{Letter} we extend and complement the work of  Hahn et al. (2007, 2010)  and \cite{Aragon-Calvoetal2007} in two ways. First we examining the orientation of the halo with respect to the velocity shear tensor instead of the tidal tensor. This method allows us to obtain much higher spatial resolution for our web classification. Secondly, we look at the orientation with respect to the cosmic web of substructure angular momentum, in a bid to understand how the internal dynamics of haloes and the substructures resident within them are effected by and reflective of the large scale cosmic web.

\section{Methods}
\label{sec:methods}

\begin{figure}
\includegraphics[width=20pc]{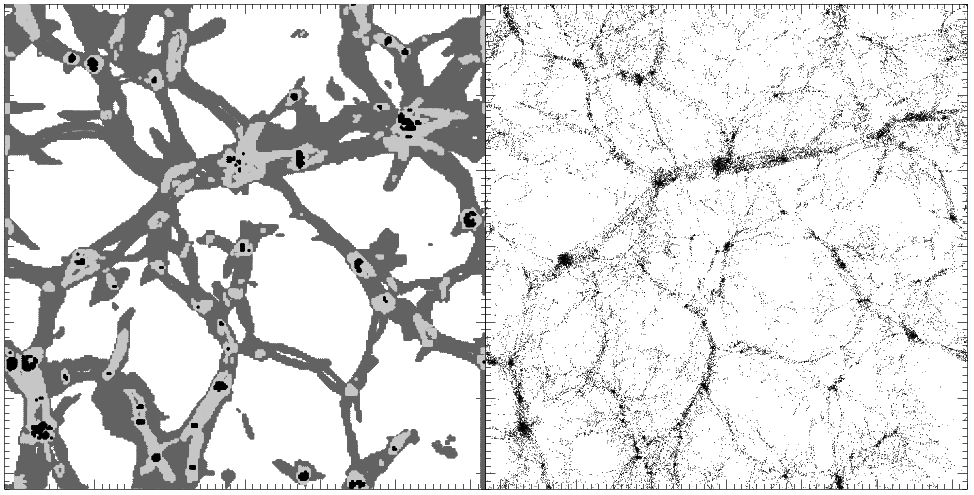}
\caption{A 0.25$h^{-1}$~Mpc slice through the simulations box of side length 64$h^{-1}$Mpc. In the left panel, we show the mesh cells associated with knots (black), filaments (light gray) and sheets (dark gray). On the right we show the distribution of DM haloes in the same simulation slice.}
\label{fig:simfig}
\end{figure}

We use a DM only $N$-body cosmological simulation run assuming the standard $\Lambda$CDM concordance cosmology \citep[e.g. WMAP5,][]{Komatsu09}. These assume a flat universe with cosmological constant density parameter $\Omega_{\Lambda}=0.72$, matter density parameter $\Omega_{\rm m}=0.28$, a Hubble constant parameterized by $H_{0}=100h$~km~s$^{-1}$~Mpc$^{-1}$ (with $h=0.7$), a spectral index of primordial density fluctuations given by $n_{s}=0.96$, and mass fluctuations given by $\sigma_{8}=0.817$.

The simulations span a box of side length $64h^{-1}$Mpc with 1024$^{3}$ particles, achieving a mass resolution of $\sim 1.89\times 10^{7}h^{-1}M_{\odot}$ and a spatial resolution of 1$h^{-1}$kpc. The publicly available halo finder AHF \citep{Knollman09} is run on the particle distribution to obtain a halo catalogue. AHF identifies haloes and subhaloes in the simulation by searching the particle distribution for local density by maxima and checking that particles within the virial radius are gravitationally bound to the host structure. Substructures are identified as haloes whose center is located within the virial radius of a more massive parent halo.

\begin{figure}
\includegraphics[width=22pc]{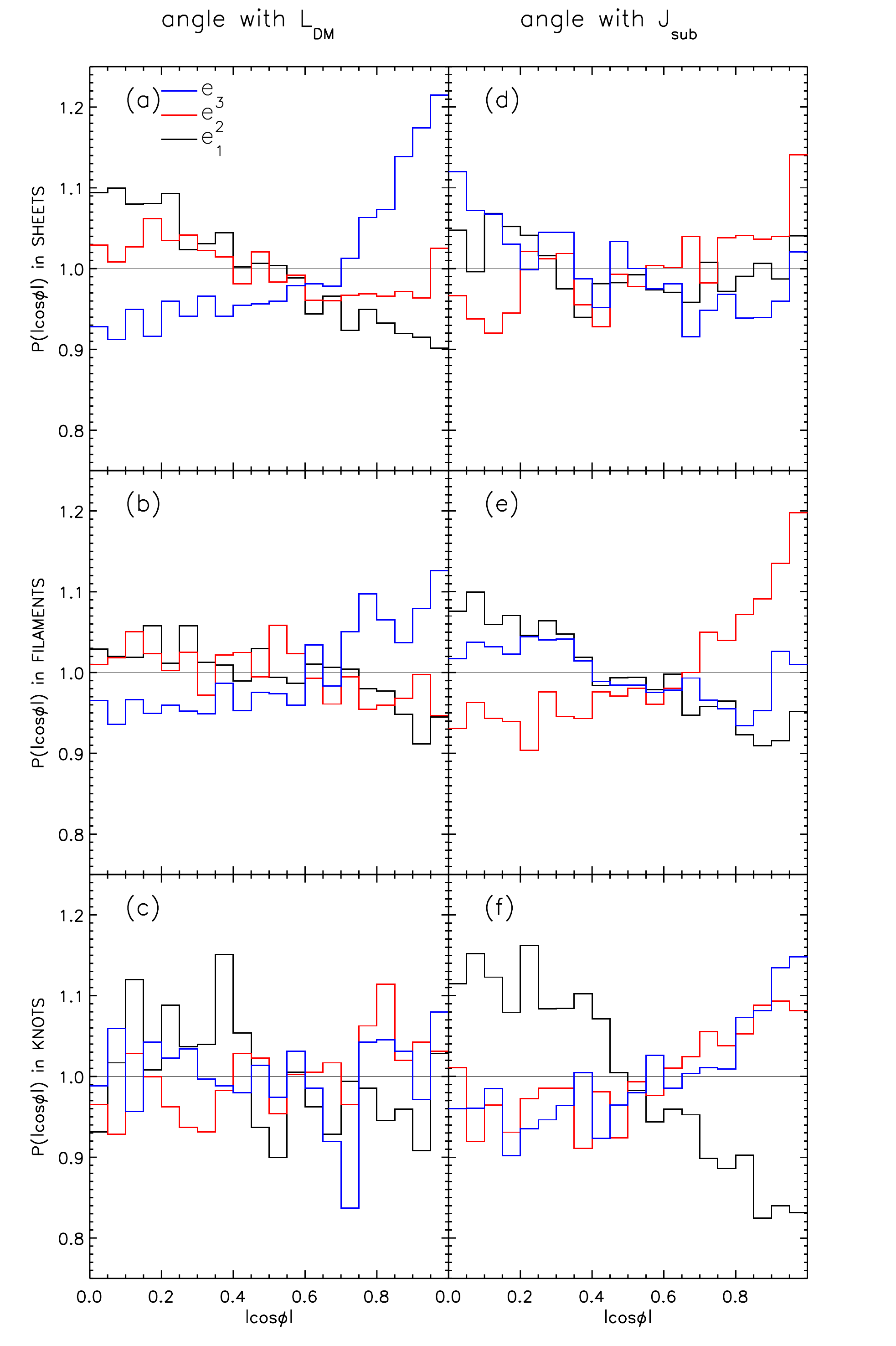}
\caption{Normalized histograms of the cosine of the angle between the spin axis of the halo ($L_{\rm DM}$, left), the orbital angular momentum of substructures ($J_{\rm sub}$ right) and the principal axes of the cosmic web in sheets (top), filaments (middle) and in knots (bottom). We show the angle formed with the long ($\hat{e}_{1}$, black), intermediate ($\hat{e}_{2}$, red) and short ($\hat{e}_{3}$, blue) axes of the cosmic web. A uniform distribution of angles is shown by the flat line at $P(|\cos\phi|)=1$.}
\label{fig:histfig}
\end{figure}

A grid based velocity field is constructed by the ``clouds in cells'' (CIC) algorithm. The velocity of each cell is computed by summing the momentum of all the particles in each cell and then dividing by the total mass in that cell. A normalized shear tensor is defined at each grid cell point by $\Sigma_{\alpha\beta}=-\frac{1}{2H_{0}}\big(\frac{\partial v_{\alpha}}{\partial r_{\beta}}+\frac{\partial v_{\beta}}{\partial r_{\alpha}}\big)$, where $\alpha$ and $\beta$ are the x, y, and z components. The $H_{0}$ normalization is used to make the tensor dimensionless and the minus sign is introduced to make the positive eigenvectors correspond to a converging flow. The shear tensor is diagonalized at each grid point to obtain the eigenvalues $\lambda_{i}$ and the corresponding eigenvectors $\hat{e}_{i}$, (where $i=1,2,3$ and $\lambda_{1}>\lambda_{2}>\lambda_{3}$). The web classification at each grid cell is done by counting how many eigenvalues are above a threshold $\lambda_{\rm th}=0.07$, (chosen in Hoffman et al 2011 to visually reproduce the cosmic web). If 0, 1, 2, or 3 eigenvalues are above $\lambda_{\rm th}$ we classify the grid cell as a void, sheet, filament or knot respectively. The CIC velocity  field is then gridded into $256^{3}$ cells, each with a length of 0.25~$h^{-1}$Mpc and is then Gaussian smoothed with a kernel of 0.25~$h^{-1}$Mpc so as to suppress numerical artifacts and spurious preferred directions introduced by the cartesian grid. Note that the CIC algorithm inherently smoothes the discrete particle distribution across eight cells, two for each of the three dimensions. This intrinsic smoothing allows us to use a gaussian smoothing of 1 grid cell while still suppressing artificial grid effects.

Note that the number of positive eigenvalues corresponds to the number of eigenvectors along which mass is moving ``inwards'' or collapsing, while the number of negative eigenvalues indicates expansion. In this work we assume that if  $0<\lambda_{i}<\lambda_{\rm th}$ the eigenvector is collapsing so slowly as to be considered expanding. Thus knots and voids are collapsing and expanding in all three directions respectively. Filaments, expand along their long axis, $\hat{e}_{3}$, while collapsing along $\hat{e}_{1}$ and $\hat{e}_{2} $ while sheets collapse along their short axis $\hat{e}_{1}$, and expand in the plane defined by $\hat{e}_{2}$ and $\hat{e}_{3}$. 

We focus on the orientation of halo angular momentum ($L_{\rm DM}$) and the orbital angular momentum of subhaloes ($J_{\rm sub}$) with respect to the cosmic web. Often the word ``orientation'' is used to mean the orientation of these vectors.  For the study of $L_{\rm DM}$ only virialized haloes more massive that $10^{9.5}M_{\odot}$ are considered; this returns $\sim 10^{6}$ haloes. Of these, $\sim2.7\times10^{4}$ haloes contain $\sim1.2\times10^{5}$ substructures more massive than 30 particles, that are then used in the study of $J_{\rm sub}$. Only subhaloes with $>$30 particles are used here \citep[\textsc{AHF} returns converged results for haloes above this limit, see][]{2011MNRAS.415.2293K}, but note that increasing or decreasing our subhalo resolution to 20 or 200 particles, has no effect on our results.

The alignment of $L_{\rm DM}$ and $J_{\rm sub}$ is measured with respect to the eigenvectors of the shear tensor. To each host halo a cosmic web classification can be assigned by finding within which grid cell the halo is located. It is the orientation of halo properties with respect to that grid cell's orthonormal basis of the V-web that is examined here. 

For the purpose of this \textit{Letter} we examine only haloes that are found at $z=0$ within sheets, filaments and knots. We ignore the orientation of haloes with respect to voids for two reasons (see table~\ref{table:web} and discussion in section~\ref{sec:results} for details): the number density of haloes in voids is exceedingly small when compared to the other web classifications and the haloes in voids tend to be low mass - little of the mass bound to haloes at $z=0$ is in voids.

\section{Results}
\label{sec:results}
Fig.~\ref{fig:simfig} shows a 0.25~$h^{-1}$~Mpc slice through the centre of our simulation box. Since each point (grid cell) in space is given a web classification, we can visualize the cosmic web by dividing it into its constituent components. Fig.~\ref{fig:simfig} reveals the nature of the cosmic web: knots are fragmented over dense clumps in the puffier filaments, which themselves inhabit the centers of even less dense sheets. The filamentary appearance of sheets in Fig.~\ref{fig:simfig} is a result of the planar cut of the two dimensional manifold. Although it appears that sheets dominate the environment of haloes this is only due to their large volume. In Table~\ref{table:web} we present the properties of the cosmic web in terms of various filling factors. 

This table reveals the nature of the halo distribution - for example, although voids occupy 69\% of the simulation volume, only 21\% of the haloes are found in voids. Furthermore, these haloes tend to be small, making up just 5\% of the mass bound to haloes. Finally since the volume occupied by voids is so large, the number density of haloes here is 0.3 times the mean while the mass density is  just 0.08 times the mean mass density of the haloes in the universe. Conversely for knots, we see that although they occupy just 0.5\% of the simulation's volume, they contain a quarter of the mass bound to haloes and are 72 times denser then the mean density.

\begin{table}
\begin{center}
 \begin{tabular}{l l l l l}
&Knots & Filaments & Sheets & Voids\\
   \hline
   \hline
$V_{ff}$  &  0.5\% & 4.5\% & 26\% & 69\% \\
$N_{f}$  &  6\% & 28\% & 45\% & 21\% \\
$M_{f}$  &  25\% & 44\% & 26\% & 5\% \\
$N$  &  17 & 6.1 & 1.7 & 0.3 \\
$\rho$  &  72 & 9.4 & 0.96 & 0.08 \\
 \end{tabular}
 \end{center}
\caption{Properties of haloes and the cosmic web. 
We present: the volume filling fraction $V_{ff}$ (the fraction of the total volume occupied by each web type), the number fraction $N_{f}$ (the fraction of haloes assigned to a given web type), the halo mass fraction $M_{f}$ (the fraction of all halo mass, in haloes of each web type), the number density of haloes $N$ (the number of haloes in a given web type divided by the volume occupied by that web type, normalized by the mean number density of haloes), and the mass density of haloes $\rho$ (the total mass of haloes in each web type divided by the volume occupied by that web type, normalized by the mean halo mass density).}
\label{table:web}
\end{table}

Fig.~\ref{fig:histfig} shows the normalized (differential) distribution of the angles formed between the principal axes of the cosmic web with the halo and satellite orbital angular momenta $L_{\rm DM}$ and $J_{\rm sub}$ for haloes in a variety of different cosmic web environments. We performed Kolmogorov-Smirnoff (KS) tests to check the null hypothesis that the distributions are consistent with being drawn from a random (uniform) distribution. With the exception of the Knots, the null hypothesis can be ruled out at high ($>$99.99\%) confidence levels. We examine these panels in detail below.

In Fig.~\ref{fig:histfig}a-c, we show the alignment of $L_{\rm DM}$ with the three principle axes of the cosmic web. In sheets (Fig.~\ref{fig:histfig}a) we see a clear alignment with $\hat{e}_{3}$ and a (slightly weaker) perpendicular alignment with $\hat{e}_{1}$ and  $\hat{e}_{2}$. This indicates that a halo's spin axis lies in the sheet along the axis that is collapsing slowest ($\hat{e}_{3}$) and perpendicular to the sheet's short (normal) axis ($\hat{e}_{1}$). KS tests rule out uniformity at a high confidence level for all three axes. When we examine the orientation of $L_{\rm DM}$ in filaments (Fig.~\ref{fig:histfig}b) we find a general weakening of the alignments found in sheets, but the same over all picture prevails - the halo spin axis points along the long axis of the filament ($\hat{e}_{3}$). Again, KS tests rule out uniformity at a high confidence level. The alignment weakens further in knots (Fig.~\ref{fig:histfig}c) where the distribution of $L_{\rm DM}$ is fully consistent with a random orientation with respect to all three principal axes (KS probabilities of $\sim 20\%$).

Haloes that live in sheets are aligned perpendicular to the sheet's short axis, while filament haloes are aligned parallel to the filamentary axis, a result consistent with previous work \citep[e.g.][]{Hahnetal2007a, Aragon-Calvoetal2007}. In sheets, this is because the short axis is the axis of collapse, hence the axis from which matter is being gravitationally pulled along. If the sheet is squeezed along this axis, it follows that the angular momentum will be perpendicular to $\hat{e}_{1}$. Yet this interpretation weakens in filaments (as there are two ``short axes'') and disappears in knots. As haloes are drawn into filaments and then knots, their angular momentum becomes randomized, most likely due to the highly non linear chaotic motions present in denser environments. 

Fig.~\ref{fig:histfig}d-f shows the alignment of $J_{\rm sub}$, the orbital angular momentum of substructures with the cosmic web. It is immediately clear that  that $J_{\rm sub}$ is always aligned parallel to the intermediate axis of the cosmic web - this alignment is strongest in filaments yet seen in all web classifications at significant KS confidence levels and is indicative of similar findings by \cite{NavarroAbadiSteinmetz04}.  In both sheets and filaments however, there is additionally a (slightly weaker) perpendicular alignment with the small and long axes of the web. This indicates that substructures are accreted (and are thus moving) in a plane fairly well defined by $\hat{e}_{1}$ (the sheet's short axis) and $\hat{e}_{3}$ (the sheet/filaments long axis).
 
Remarkably the perpendicular alignment with $\hat{e}_{3}$ radically and significantly flips in knots (Fig.~\ref{fig:histfig}i). This indicates that although substructures are accreted in the $\hat{e}_{1}$-$\hat{e}_{3}$ plane, by the time they land in knots the plane has nutated such that $J_{\rm sat}$ is perpendicular to just $\hat{e}_{1}$, the main axis of the knot. The plane of accretion in filaments is well defined - the $\hat{e}_{1}$-$\hat{e}_{3}$, while in knots this plane is allowed to nutate about $\hat{e}_{2}$.

It is important to note that the significant  perpendicular alignment of $J_{\rm sub}$ with $\hat{e}_{1}$ in knots is strong evidence that the cosmic web is well defined in knots. This indicates that the uniform distribution of angles seen between the cosmic web and $L_{\rm DM}$ in knots (Fig.~\ref{fig:histfig}c) is indeed due to a randomization of $L_{\rm DM}$.

We close this section by noting that we also examined the orientation of substructure position with the cosmic web. Since the position vector is by construction perpendicular to $J_{\rm sub}$ we find a mirror image panels (d)-(e), in otherwords, the positions of subhaloes tend to align with $\hat{e}_{1}$ and $\hat{e}_{3}$ in sheets and filaments and just with $\hat{e}_{1}$ in knots. 

\section{Discussion and Conclusions}
\label{sec:conclusion}

We have examined the orientation of DM spin axis ($L_{\rm DM}$) and the orbital angular momentum of substructures ($J_{\rm sub}$) with respect to the cosmic web in sheets, filaments and voids. Our main findings are 
\begin{itemize}
\item Substructures orbit such that their angular momenta tend to be aligned with the intermediate axis of the web structure they are in. If the second axis of the web is parallel to the intermediate axis of the material that collapsed to form the halo at turnaround, then our result is consistent with tidal torque theory \citep[e.g.][]{NavarroAbadiSteinmetz04}. This occurs because the velocity shear is closely related to the initial conditions.
\item In sheets and filaments $J_{\rm sub}$ tends to be perpendicular to the short ($\hat{e}_{1}$) \textit{and} long ($\hat{e}_{3}$) axis of the cosmic web, while in knots it is perpendicular \textit{only} to the long axis. This indicates that in sheets and filaments, substructures move in just one plane (the $\hat{e}_{1}$-$\hat{e}_{3}$) while in knots the plane of their orbit is allowed any configuration parallel to the long axis, $\hat{e}_{1}$. This transition is most likely due to the randomization of orbits in the highly non linear halo interiors.
\item The spin axis of a DM halo tends to align with the short axis of the cosmic web in sheets and in filaments, but is consistent with a random orientation in haloes associated to knots. Since in sheets the spin axis lies in the plane, and assuming that in sheets the long and intermediate axes are degenerate, this result is also consistent with \cite{NavarroAbadiSteinmetz04} (given that there is some correlation between galaxy and halo spin).
\end{itemize}

Our results indicate that the large scale velocity shear tensor is still a determining factor in the orientation of the angular momenta of substructures. Although \cite{Knebeetal2004} already claimed that the position of a subhalo at apocenter recalled its filamentary accretion even after several orbits, this link was not quantitatively established by them. This result is also consistent with a number of recent studies \citep{Libeskindetal2011a,Vera-Ciroetal2011} that have argued that a memory of the large scale environment is imprinted on $z=0$ subhalo orbits and DM haloes. Yet  none of these studies were able to draw a direct parallel between the large scales and satellite orbits  at $z=0$, instead they relied on tying $z=0$ properties of satellite galaxies to the merger history of the host halo. Indeed our result is surprising since we do no such tracking back in time. Instead our result indicates that the large scale velocity field is reflected in the orbits of subhaloes at $z=0$.

Since a subhaloes $z=0$ position is perpendicular to its angular momentum, our findings help understand the relationship between the anisotropic $z=0$ spatial distribution of satellite galaxy in simulations \citep{Knebeetal2004,Libeskindetal2005,Zentneretal2005, Libeskindetal2007} and observations \citep[e.g.][among others]{SalesLambas2004,KroupaTheisBoily2005} with their apparent coherent motion \citep{MetzKroupaLibeskind2008,MetzKroupaJerjen2009,Libeskindetal2009,Deasonetal2011}. Subhaloes are assumed to host luminous satellite galaxies that do not enter the halo randomly but in some correlated way \citep[e.g.][]{Tormen1997,LiHelmi2008,Knebeetal2011}. Our findings indicate that the orientation of the non-randomly accreted subhaloes is determined by the cosmic web at $z=0$, thus connecting the large scale sheets, filaments and knots with the sub-virial radius scale of subhalo orbits. In summary, The non-random infall points of subhaloes is both reflective of the large scale structure and  frozen into the $z=0$ angular momentum of a subhalo, while, due to long dynamical times, the $z=0$ positions of satellites are susceptible to randomization processes.

The strength of the alignment is likely to depend on a number of factors - mass of halo, redshift, and how old a given parent halo is. A comprehensive study of the nature of haloes in the cosmic web is forthcoming (Metuki et al in preparation).

\section*{Acknowledgments}
This work was supported by the Deutsche Forschungs Gemeinschaft, the Ministerio de Ciencia e Innovacion in Spain, the Ramon y Cajal program, (AYA 2009-13875-C03-02, AYA2009-12792-C03-03, CAM S2009/ESP-1496, FPA2009-08958) and through Consolider-Ingenio  SyeC. The simuations were carried out at the Leibniz Rechenzentrum (LRZ) and the Barcelona Supercomputing Center (BSC).

\bibliographystyle{mn2e}  
\bibliography{ref}
\end{document}